\documentclass{webofc}

\usepackage[varg]{txfonts}   
\usepackage{hyperref}
\usepackage{url}
\usepackage{enumitem,kantlipsum}
\hypersetup{colorlinks=true,citecolor=blue,urlcolor=blue,linkcolor=blue}
\begin{document}
\title{Using artificial neural networks in searches for Lorentz invariance violation}

\author{\firstname{Tomislav} \lastname{Terzić}\inst{1}\fnsep\thanks{\email{tterzic@uniri.hr}}}

\institute{University of Rijeka, Faculty of Physics, Rijeka 51000, Croatia}

\abstract{Lorentz invariance violation (LIV) in gamma rays can have multiple consequences, such as energy-dependent photon group velocity, photon instability, vacuum birefringence, and modified electromagnetic interaction. Depending on how LIV is introduced, several of these effects can occur simultaneously. Nevertheless, in experimental tests of LIV, each effect is tested separately and independently. For the first time, we are attempting to test for two effects in a single analysis: modified gamma-ray absorption and energy-dependent photon group velocity. In doing so, we are using artificial neural networks. In this contribution, we discuss our experiences with using machine learning for this purpose and present our very first results.
}
\maketitle
\section{Introduction and motivation}
\label{sec:intro}

Some theoretical approaches to quantum gravity (QG) propose that at extremely high energies space-time may no longer be smooth and continuous, and small Lorentz invariance violation (LIV) might arise. Measuring such deviations could steer the search for the theory of QG.
We focus on searching for effects of LIV using very high energy gamma rays (VHE, $E > 100$\,GeV). 
LIV is usually modelled as a modified photon dispersion relation (MDR)

\begin{equation}
\label{eq:MDR}
    E^{2}\simeq p^{2}\times\left[1+\sum_{n=1}^{\infty} S_n \left(\frac{E}{E_{\mathrm{QG,}n}}\right)^{n}\right],
\end{equation}
where $E$ and $p$ are gamma-ray energy and momentum, respectively, $c$ is the standard speed of light, $S_n$ is a prefactor which can take values $\pm1$, and $E_\mathrm{QG}$ is the QG energy scale, typically expected to be on the order of Planck energy $E_\mathrm{P} \approx 1.2 \times 10^{19}$\,GeV. While MDR is not a direct consequence of any particular QG model, it is a simple way to parametrize non-standard effects, and the usual starting point in searches for LIV effects with gamma rays. 

Consequences of MDR include energy dependent photon group velocity, modified reaction thresholds and cross sections, vacuum birefringence, etc. All of these have been tested, but so far, no trace of LIV has been found. Depending on the underlying theory, some of these effects can coexist. 
For example, quantum electrodynamics (QED) with LIV operators of dimension up to six in the Lagrangian, assuming preserved gauge invariance and CPT and P invariance, will lead to energy-dependent photon group velocity and anomalous QED interactions~\cite{Rubtsov:2012kb, PhysRevD.110.063035}. Therefore, VHE gamma rays will experience a time of flight from source to detector to be energy dependent, while their emission, interaction with the background radiation, and detection will be affected by LIV. 
Nevertheless, all of the experimental studies test only one of the effects independently. To date, only one study tested two effects on the same data set, alas again treating them mutually independent~\cite{HESS:2019rhe}.

In this work, we are reporting on the first attempt to test two coexisting effects of LIV in a single analysis. Namely, we consider the gamma-ray arrival time delay and the probability of survival to the detector without being absorbed on the extragalactic background light (EBL). The LIV effect of the former is given as~\cite{DistancePiran}
\begin{equation}\label{eq:delay}
    \Delta t_n \cong - S_n  \frac{n+1}{2} \frac{E^{n}}{ E_{\mathrm{QG}, n}^{n}} \kappa_n (z_{\mathrm{s}}),\quad \kappa_n (z_{\mathrm{s}}) = \frac{1}{H_0}\int_0^{z_{\mathrm{s}} }\frac{(1+z)^n}{\sqrt{\Omega_\Lambda + \Omega_m(1+z)^3}}\;\mathrm{d}z.
\end{equation}
For the standard cosmological parameters of the $\Lambda$CDM cosmology model, we use $H_0 = 70\,\mathrm{km\,Mpc}^{-1}\,\mathrm{s}^{-1}$, $(\Omega_\Lambda, \Omega_m) = (0.7, 0.3)$.
The survival probability $P(E, z_s) = \exp\left(-\tau(E, z_{\mathrm{s}})\right),$
depends on the opacity
\begin{equation}\label{eq:opa}
    \tau(E,z_s) = \int_{0}^{z_{\mathrm{s}}}dz\,\frac{dl}{dz}\int_{-1}^{1} d\cos\theta \left(\frac{1-\cos\theta}{2}\right) \int_{\omega_\text{th}(E,\theta)}^\infty d\omega \; n(\omega,z) \,\sigma(E(1+z),\omega,\theta), 
\end{equation}
where $E$ is the gamma-ray energy at detection, $z_s$ the source redshift, $n$ the EBL spectral density, and $\omega_\text{th}$ the reaction threshold of the soft photon. In this case, LIV affects both $\omega_\text{th}$ and the cross section $\sigma(E(1+z),\omega,\theta)$. It is important to note that the expressions for cross section usually used in the literature are approximations, and not good ones. We follow the recommendation from~\cite{PhysRevD.110.063035}, which offers the most complete calculation up to date.
For simplicity, we restrict ourselves to the case $n=2$, i.e. only second order modification in Eq.~(\ref{eq:MDR}) is considered, and $S_2 = -1$.

While such analysis should be possible using standard analysis tools, such as likelihood maximisation (e.g.~\cite{Martinez:2008ki, Abe_2024, HESS:2019rhe, Terzic:2021rlx}), adding more stages of gamma-ray life in the analysis, such as inverse-Compton scattering or pion decay at the emission, or extensive air shower (EAS) development as a part of the detection process, these approaches might prove unmanageable. Therefore, we are attempting to use artificial neural networks (ANN) to search for LIV.

\section{Artificial neural network training}
\label{sec:DataSet}
We use the Mrk~501 flare from 2014, observed by the H.E.S.S. Collaboration~\cite{HESS:2019rhe} as a sandbox.
Both the spectrum and light curve have rather simple functional forms. Moreover, since this is the only study so far in which the two LIV effects were tested on the same sample (although independently), it should allow us to compare the sensitivity of our approach. 
We generated a training sample modelled on the Mrk~501 data set through the following steps:
\begin{enumerate}[wide, labelwidth=!, labelindent=0pt]
	\item We generated $10^6$ ordered pairs $(t_e, E_e)$, where $t_e$ represents the emission time, and $E_e$ the gamma-ray energy at emission. Temporal distribution of gamma rays was a double Gaussian with the following parameters $(A_1, \mu_1, \sigma_1) = (80.5, 2361, 2153)$, $(A_2, \mu_2, \sigma_2) = (60.5, 6564, 676)$, where $A_i$, $\mu_i$, $\sigma_i$ are the prefactor, mean, and width of each Gausian, respectively. The time interval is $[0, 7200]$\,s. The spectral distribution of events is a power-law with slope $\alpha = -2.03$ for energies between ~35\,GeV and ~50\,TeV.
	\item We calculate the detection time $t_\mathrm{d} = t_\mathrm{e} + \Delta t_2$, the detected energy $E_\mathrm{d} = E_\mathrm{e} / (1+z_\mathrm{s})$, and the survival probability for every event. The time delay and survival probability are calculated for $E_\mathrm{QG} \rightarrow\infty$, which corresponds to Lorentz invariant case, and 209 different values of $E_\mathrm{QG}$, logarithmically distributed on the $[10^9, 10^{30}]$\,eV interval.
	\item The training sample is created by randomly picking from the pool of $10^6$ events. For each event, the survival probability is compared to a random number $p_i \in [0, 1]$. All the events are considered to be emitted, but only the ones with $P(E, z_s, E_\mathrm{QG})>p_i$\footnote{Note that now the survival probability is also a function of $E_\mathrm{QG}$.} are detected. The size of the detected sample is 2000. The distributions of events in both emitted and detected samples are fitted in time with a double Gaussian, and in energy with a power law. The process is repeated 100 times for each $E_\mathrm{QG}$ value, generating in total 21,000 training samples.
\end{enumerate}
The training sample prepared in such a way assumes no source-intrinsic correlation between gamma-ray energy and emission time. In addition, we assumed a perfect detector.

The ANN consists of the input layer with 8 nods, one for each parameter describing the detected sample: 6 light curve parameters, and 2 spectral parameters. There are three hidden layers of 1000 nods each. The output layer consists of 9 nods: 8 for the light curve and spectral parameters of the emitted sample, and one representing LIV. All input and output parameters are mapped to a $[0, 1]$ interval.
The training is performed with 2500 iterations.

\section{Results}
\label{sec:results}
\begin{figure}[h]
\centering
\includegraphics[width=4cm,clip]{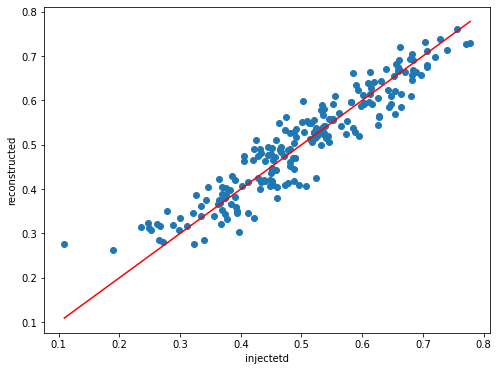}~~
\includegraphics[width=4cm,clip]{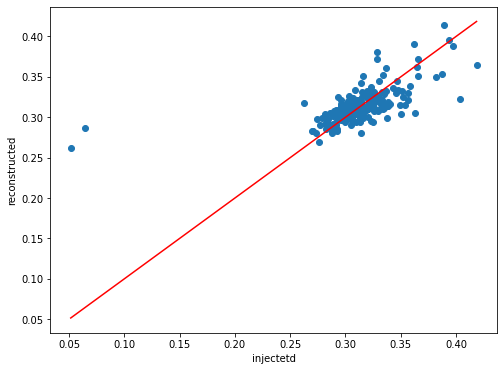}~~
\includegraphics[width=4cm,clip]{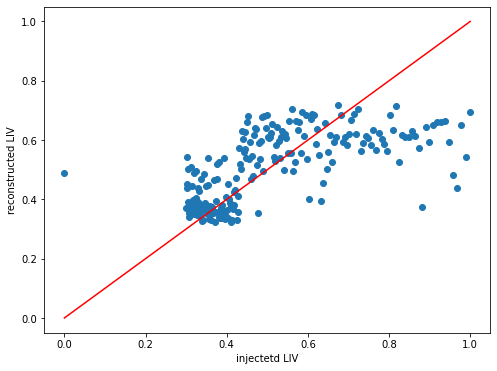}
\caption{Comparison of injected and reconstructed parameter values. \textit{Left:} Randomly chosen parameter from the light curve or spectral fit, with homogeneous distribution of values. \textit{Centre:} Randomly chosen parameter, with inhomogeneous distribution. \textit{Right:} LIV parameter. The red solid line shows the ideal reconstruction.}
\label{fig:fig1}    
\end{figure}

To test the ANN, we feed it with parameters from the detected sample, and compare the parameters’ values in the emitted sample with the values reconstructed by the ANN. The results are shown in Figure~\ref{fig:fig1}.
There is a strong correlation between true and reconstructed values of parameters that appear both in the emitted and detected sample, and whose values are homogeneously distributed over the $[0, 1]$ interval. 
An example in the left plot of Figure~\ref{fig:fig1}, has the Pearson correlation coefficient $(r, p) = (0.94, 10^{-98})$. 
On the other hand, when the $[0, 1]$ interval is not well covered, the correlation is not as strong. This can be seen in the central plot of Figure~\ref{fig:fig1}, where, in addition to inhomogeneous distribution of the points, there are some outliers. In this case the correlation is $(r, p) = (0.63, 10^{-25})$. While $r=0.63$ is considered to be a strong correlation, we aim for values much closer to unity for good reconstructive power. A similar situation happens with the LIV parameter (right plot of Figure~\ref{fig:fig1}), where $(r, p) = (0.64, 10^{-25})$. Just as in the central plot, the coverage can be optimised. This part is somewhat tricky because the null hypothesis corresponds to an infinite value of $E_\mathrm{QG}$, and special attention should be paid to seamlessly connecting this case with finite values of $E_\mathrm{QG}$.

\section{Discussion and outlook}
\label{sec:discussion}

In this contribution we presented the first attempt at simultaneous testing of two coexisting LIV effects in the VHE gamma-ray data. We used a simple ANN structure, which successfully reconstructed some of the parameters which were present both in input and output layers of training samples. The parameter representing LIV was present in the output layer only, and in this case, the reconstruction power is substantially weaker. There is obviously room for improvement. Firstly, by more adequate mapping of the LIV parameter to the $[0, 1]$ interval and parameter space coverage, and secondly, through ANN hyperparameter optimization and different choices of training samples. These constitute our short-term plans for the continuation of this research.

Our mid-term plans include investigating different approaches; e.g. using individual events or bins in the input layer, instead of fit parameters. Furthermore, we will introduce realistic detectors with non-unit acceptance and finite energy resolution). That will be followed by an estimate of the sensitivity to systematic effects and comparison to the sensitivity to other analysis methods, such as likelihood.

On the long-term, we intend to combine multiple observations, introduce additional LIV effects (e.g. EAS development), introduce additional free parameters representing source intrinsic correlation, cosmology, background models, etc.

\begin{acknowledgement}
The author acknowledges the networking support by COST actions CA18108 (QG-MM, \url{https://qg-mm.unizar.es/}) and CA23130 (BridgeQG, \url{https://web.infn.it/BridgeQG/}), and the financial support from the University of Rijeka through project uniri-iskusni-prirod-23-24 and the Croatian Science Foundation (HrZZ) Project IP-2022-10-4595.
We thank the Institute for Fundamental Physics of the Universe (IFPU, \url{https://www.ifpu.it/}) for hosting the workshop ``Astrophysical searches for quantum-gravity-induced time delays'' where ideas important to this work were developed.
\end{acknowledgement}

\bibliography{main}

\end{document}